\documentclass[preprint,aps,prc,showpacs,nofootinbib]{revtex4}
\usepackage{epsfig}
\usepackage{graphicx,amsmath}

\newcommand\ba{\begin{eqnarray}}
\newcommand\ea{\end{eqnarray}}
\newcommand\be{\begin{equation}}
\newcommand\ee{\end{equation}}
\newcommand\nn{\nonumber}

\begin{document}

\title{QED processes in peripheral kinematics at polarized
photon-photon and photon-electron colliders}

\author{S.~Bakmaev}
\affiliation{\it JINR-BLTP, 141980 Dubna, Moscow region, Russian Federation}

\author{E.~Barto\v s}
\affiliation{Institute of Physics SAS, 84511 Bratislava,
Slovakia}

\author{E.~A.~Kuraev}
\affiliation{\it JINR-BLTP, 141980 Dubna, Moscow region, Russian Federation}
\date{\today}

\author{M.~V.~Galynskii}
\affiliation{Institute of Physics BAS, Minsk, 220072, Belarus}


\begin{abstract}
The calibration QED process cross sections for experiments on planned
electron-photon and photon-photon colliders for detecting the small angles scattered particles
 are calculated. These processes describe the creation of two jets moving sufficiently
close to the beam axis directions. The jets containing two and
three particles including charged leptons, photons and
pseudoscalar mesons are considered explicitly. Considering the
pair production subprocesses we take into account both
bremsstrahlung and double photon mechanisms. The obtained results
are suitable for further numerical calculations.
\end{abstract}
\maketitle

\subsection*{Introduction}
QED processes of the type $2\to 3, 4, 5, 6$ at colliders of high energies have  attracted
both theoretical and experimental attention during the last four
decades. Accelerators with high-energy colliding $e^+e^-$, $\gamma
e$, $\gamma\gamma$ and $\mu^+ \mu^-$ beams are now widely used or
designed to study fundamental interactions \cite{NLC}. Some
processes of quantum electrodynamics (QED) might play an important
role at these colliders, especially those inelastic processes whose
cross section does not drop with increasing energy. The
planned colliders will be able to work with polarized particles,
so these QED processes are required to be described in more
detail, including the calculation of cross sections with definite
helicities of the initial particles -- leptons ({\it l}\,=\,e or
$\mu$) and photons $\gamma$. These reactions have the form of a
two-jet process with the exchange of a virtual photon $\gamma^*$
in the t-channel (see Fig.1).

A lot of attention to the calculation of helicity amplitudes
of QED processes at high energy colliders was paid in the literature
(see \cite{C} and references therein).
Keeping in mind the physical programs at planned $\gamma\gamma$
and lepton $\gamma$ colliders, a precise knowledge of a set of
calibration and monitoring processes is needed. The
calibration processes are the QED processes with sufficiently large
cross sections and clear signatures for detection. Rather a rich
physics can be investigated in peripheral processes such as heavy leptons and
mesons (scalar and pseudoscalar) creation, where the relevant QED
monitoring processes must be measured.

Let us remind the general features of peripheral processes
namely the important fact of their nondecreasing cross sections in
the limit of high total energies $\sqrt{s}$ in the center of mass
frame of the initial particles. The possibility of
measuring the jets containing two or three particles can be
relevant. This is a motivation of our paper.

It is organized in the following a way.

In Section 1, the kinematics  of peripheral processes is briefly described.

In Section 2, the impact factors describing the conversion of
initial photon, to the pair of charged particles (fermions or
spinless mesons)
%with or without an additional hard photon
are calculated.

In Sections 3, 4, and 5 a similar calculation is made for the initial
polarized electron and photon, in particular subprocesses such
as the single and the double Compton process, and the processes of
pair creation are considered.

As well as the helicity amplitudes for subprocesses of type $2\to
3$ have in general complicated form we do not put explicit
expressions for the corresponding cross sections indicating only
the strategy to obtain it.

\begin{figure}
\begin{center}
\includegraphics[scale=.8]{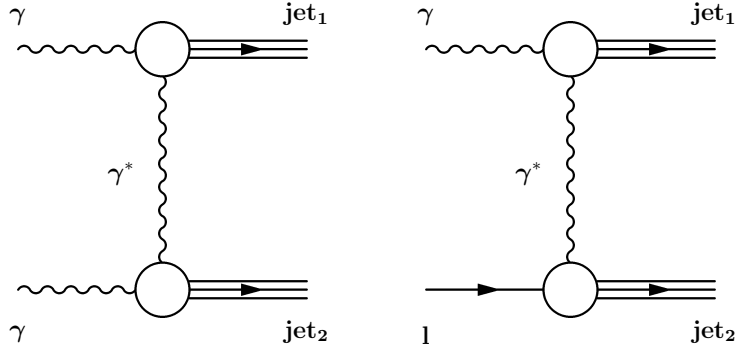}
\end{center}
\caption{The processes $\gamma\gamma$, $\gamma l$ ($l=e,\mu$) with
the exchange of a virtual photon $\gamma^*$ in the t-channel.}
\end{figure}

\begin{figure}
\begin{center}
\includegraphics[scale=1]{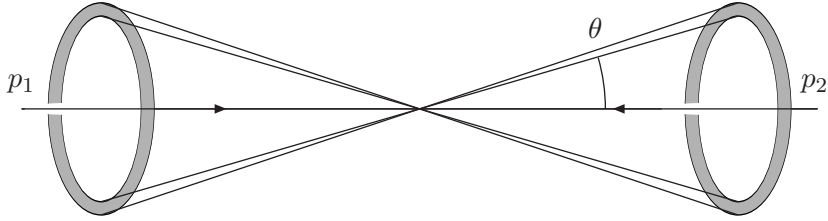}
\end{center}
\caption{The scheme of collision of initial beams with detection
of two jets moving in the cones within the angles $\theta$.}
\end{figure}

\begin{figure}
\begin{center}
\includegraphics[scale=.9]{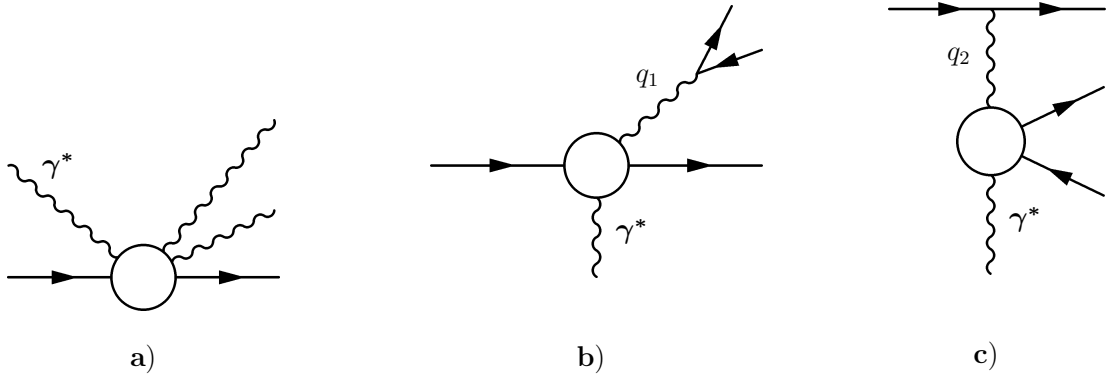}
\end{center}
\caption{Feynman diagrams describing a) the subprocess
$\gamma^\ast e^- \to \gamma\gamma e^-$
 and pair production $\gamma^\ast e \to e a \bar{a}$ subprocess by the
 bremsstrahlung b) and double photon c) mechanisms.}
\end{figure}

%The jet kinematics in QED a high-energy reaction in which the
%outgoing particles(leptons and photons) are produced within a
%small but enough to provide the detection of jet particles cones
%relative to the propagation axis of their respective parental incoming %particle.

\subsection{Kinematics}

Throughout the paper it is implied that the energy fractions of jet
component are positive quantities of the order of unity by magnitude
(the sum of energy fractions of each jet is unity) and the values
of transversal to the beam direction component of their 3-momenta are
much larger compared to their rest masses. So we neglect the mass
of jet particles.

The corresponding amplitudes include a large amount of Feynman diagrams (FD).
Fortunately, in the high-energy limit the number of essential FD contributing the
 "leading" approximation greatly reduces. The method used permits one to estimate
the uncertainty caused  by "nonleading" contributions which have following magnitudes of the order
 \begin{eqnarray} \frac{m^2}{s_1}\,,\quad\frac{s_1}{s}\,,\quad
\frac{s_2}{s}\,, \quad \frac{\alpha}{\pi}\ln{\frac{s}{m^2}} \end{eqnarray}
where $s_{1,2}$ - is the jet invariant mass squares compared with the terms of order unity. The last term
in (1) is caused by the absence of radiative corrections in our analysis. The angles $\theta_i$ of
particle emission to the corresponding projectile
direction of motion is assumed to be of the order (see Fig.2)
\begin{eqnarray}
\frac{m_i}{\sqrt{s}}\ll \theta_i \sim
\frac{\sqrt{s_i}}{\sqrt{s}}\ll 1, \end{eqnarray}
where $m_i$ is the typical mass of the jet particle.

In this approach we can consider initial particles (having the
4-momenta $p_1,p_2$) as a massless and use the Sudakov
parameterization of 4-momenta of any particle of the problem \begin{eqnarray}
q_i=\alpha_i p_2+\beta_i p_1+q_{i\bot}\,, \\ \nonumber
q_{i\bot}p_{1,2}=0\,,\quad q_{i\bot}^2=-\vec{q}_i^{~2}<0\,. \end{eqnarray} The
Sudakov parameters $\beta_i$ are the quantities of order of
unity for the momenta of the particles belonging to the $jet1$ and
obeying the conservation law $\sum_{jet1}\beta_i=1$, whereas the
components of the $jet1$ particle momenta along the four momentum
$p_2$ are small positive numbers which can be determined from the
on mass shell conditions of the $jet1$ particles
$q_i^2=s\alpha_i\beta_i-\vec{q}_i^{~2}=0,
\alpha_i=\vec{q}_i^{~2}/(s\beta_i)\ll 1$.

The  same is valid for the 4-momenta of the particles belonging
to the $jet2$, namely, $\alpha_j\sim 1$, $\sum_{jet2}\alpha_j=1$,
$\beta_j=\vec{q}_j^{~2}/(s\alpha_j)\ll1$.

Among the large amount of Feynman diagrams (FD), describing the
process in the lowest (Born) order of perturbation theory (PT)
(tree approximation), only ones survive (i.e., give a contribution
to the cross section which do not decrease with increasing $s$)
which have a photonic t-channel one-particle state.

It is known \cite {BFKK} that the matrix elements of the peripheral processes
have a factorized form and the cross section can be written in
terms of the so-called impact factors, each of which describe the
subprocess of interaction of the internal virtual photon with one
of the initial particles to produce a jet moving in the direction
close to this projectile momentum. So the problem can be
formulated in terms of computation of impact factors. For
processes with initial photons with definite state of polarization
described in terms of Stoke's parameters we construct the relevant
chiral matrices from bilinear combinations of chiral amplitudes.
The last step consists in the construction of differential cross
sections.

The  matrix element, which corresponds to the main ("leading") contribution, to the cross section, has the form
\begin{eqnarray}
M=iJ_1^\mu\, \frac{g_{\mu\nu}}{q^2}\,J_2^\nu,
\end{eqnarray}
where $J_1^\mu$ and $J_2^\nu$ are the currents of the
upper (associated with $jet1$) and lower blocks of the relevant
Feynman diagram,
respectively, and $g_{\mu\nu}$ is the metric tensor. The current
$ J_1^\mu$ describes the scattering of an incoming particle
of momentum $p_1$ with a
virtual photon and subsequent transition to the first jet (similar for $J_2^\nu$).
Matrix elements (4) can be written in the form (see the appendices in \cite{BFKK})
\begin{eqnarray}
M=2i\frac{s}{q^2}I_1I_2, \\ \nonumber
I_1=\frac{1}{s}J_1^\mu p_{2\mu},\quad I_2=\frac{1}{s}J_2^\nu p_{1\nu}.
\end{eqnarray}
Really, it follows from the Gribov representation of the metric tensor,
\begin{eqnarray}
g^{\mu\nu}=\frac{2}{s}(p_2^\mu p_1^\nu+p_2^\nu p_1^\mu)+g^{\mu\nu}_\bot
\approx \frac{2}{s}p_2^\mu p_1^\nu.
\end{eqnarray}

Invariant mass squares of jets can also be expressed in terms of the
Sudakov parameters of the exchanged photon
\begin{eqnarray}
q=\alpha p_2+\beta p_1+q_\bot \,, \quad
(q+p_1)^2=s_1=-\vec{q}^{~2}+s\alpha\,,  \nonumber \\\quad (-q+p_2)^2=s_2=-\vec{q}^{~2}-s\beta\,,\quad
q^2=s\alpha\beta-\vec{q}^{~2}\approx -\vec{q}^{~2}\,.
\end{eqnarray}
Here and below we mean by the symbol $\approx$ the equation  with
neglect the terms which do not contribute in the limit $s\to\infty$.

The singularity of the matrix element (5) at $\vec{q}=0$ is fictitious (excluding the elastic
scattering). Really one can see that it cancels due to the current conservation
\begin{eqnarray}
q_\mu J_1^\mu\approx (\alpha p_2+q_\bot)_\mu J_1^\mu=0\,,  \quad
p_{2\mu}J_1^\mu=\frac{s}{s\alpha}\vec{q}\vec{J}_1\,, \\
q_\nu J_2^\nu\approx (\beta p_1+q_\bot)_\nu J_2^\nu=0\,,  \quad
p_{1\nu}J_2^\nu=\frac{s}{s\beta}\vec{q}\vec{J}_2\,.
\end{eqnarray}

We arrive at the modified form of the matrix element of peripheral
process
\begin{eqnarray}
M(a(p_1,\eta_1)+b(p_2,\eta_2))\to jet_{1\lambda_1}+jet_{2\lambda_2}=
i(4\pi\alpha)^{\frac{n_1+n_2}{2}}\frac{2s}{\vec{q}^2}
m_{1\lambda_1}^{\eta_1}m_{2\lambda_2}^{\eta_2},
\end{eqnarray}
where $\eta_i,$ describe the polarization states of the
projectile $i=a, b$; $\lambda_i$ describes the polarization states of participants of its initiated jet.
 The numbers of QED vertices in the upper and lower
blocks of FD (see Fig.1) are denoted by $n_{1,2}$.

We give here two alternative forms for the matrix elements $m_{1,2}$
of the subprocesses $\gamma^*(q)+a(p_1,\eta_1)$ $\to$ $jet_{1(\lambda_1)}$ and
$\gamma^*(q)+b(p_2,\eta_2)\to jet_{2(\lambda_2)}$
\begin{eqnarray}
m_{1\lambda_1}^{\eta_1}=\frac{\vec{q}\vec{J}_{1\lambda_1}^{\eta_1}}{s_1+\vec{q}^2}\,, \\
m_{1\lambda_1}^{\eta_1}=\frac{1}{s}p_{2\mu}J_{1\lambda_1}^{\eta_1\mu}\,,
\end{eqnarray}
and the similar expressions for the lower block. We use the second
representation (12). The form (11) can be used as a check of validity of gauge
invariance, namely turning the matrix elements to zero in the limit
$\vec{q}\to 0$.

A remarkable feature of the peripheral processes -- is their
differential cross sections do not depend on the total center of mass
energy $\sqrt{s}$. To see this property, let us first rearrange the
phase volume $d\Phi$ of
the final two-jet kinematics state to a more convenient form
\begin{gather}
d\Phi=(2\pi)^4\delta^4(p_1+p_2-\sum_i p^{(1)}_i-\sum_j p^{(2)}_j)dF^{(1)}dF^{(2)}=
(2\pi)^4 d^4q\delta^4_{(1)}\delta^4_{(2)}dF^{(1)}dF^{(2)}\,, \\ \nonumber
\delta^4_{(1)}=\delta^4(p_1+q-\sum_i p^{(1)}_i)\,,\quad
\delta^4_{(2)}=\delta^4(p_2-q-\sum_j p^{(2)}_j)\,, \\ \nonumber
dF_{(1,2)}=\Pi_i\frac{d^3 p_i^{(1,2)}}{2\varepsilon_i^{(1,2)}(2\pi)^3}\;.
\end{gather}
Using Sudakov's parameterization for the transferred 4-momentum $q$
phase volume
\begin{eqnarray}
d^4q=\frac{s}{2}d\alpha d\beta d^2q_\bot=\frac{1}{2s} ds_1 ds_2 d^2q_\bot\,,
\end{eqnarray}
with $s_{1,2}$ the invariant mass squares of the jets, we put the
phase volume in the factorized form
\begin{eqnarray}
d\Phi=\frac{(2\pi)^4}{2s} d^2q_\bot d s_1
dF^{(1)}\delta^4_{(1)} d s_2dF^{(2)}\delta^4_{(2)}\;.
\end{eqnarray}

Using the modified form
of the matrix element and the phase volume for the peripheral process
cross section in the case of polarized initial particles
(photons or electrons), we have
\begin{eqnarray}
d\sigma^{\eta_1\eta_2}=\frac{\alpha^{n_1+n_2}\pi^2(4\pi)^{2+n_1+n_2} d^2 q_\bot}{(\vec{q}^2)^2}
\Phi^{\eta_1}_1(\vec{q})\Phi^{\eta_2}_2(\vec{q})\,,
\end{eqnarray}
with the impact factors $\Phi_i^{\eta_i}$ in the form
\begin{eqnarray}
\Phi_i^{\eta_i}(\vec{q}) =\int ds_i \sum_{\lambda_j}|m^{\eta_i}_{i\lambda_j}|^2
dF_i\delta^4_{(i)}\,,\quad i=1,2 .
\end{eqnarray}

The matrix elements with the definite chiral states of all particles
$m^{\eta_i}_{i(\lambda)}$ , where the subscript $(\lambda)$
denotes the set of chiral parameters of the final state,
are calculated and listed below.

In the case of initial polarized photons  the
description in terms of Stoke's parameters $\xi_{1,2,3}$,
$\xi_1^2+\xi_2^2+\xi_3^2 \leq 1$ is commonly used. The matrix element squared in r.h.s. (17)  must
be replaced by \cite{RKT}
\begin{eqnarray}
T_\gamma= {\rm Sp}({\cal M}\rho)=\frac{1}{2}{\rm Sp}\left(\begin{array}{cc}m^{++}&m^{+-}\\m^{-+}&m^{--}\end{array}\right)
\left(\begin{array}{cc}1+\xi_2&i\xi_1-\xi_3\\-i\xi_1-\xi_3&1-\xi_2\end{array}\right)\,,
\end{eqnarray}
with the spin matrix ${\cal M}$ elements
\begin{eqnarray}
m^{++}=\sum_\lambda|m^+_{(\lambda)}|^2\,,\quad m^{+-}=\sum_\lambda m^+_{(\lambda)}
(m^-_{(\lambda)})^\ast\,, \\ \nonumber
m^{--}=\sum_\lambda|m^-_{(\lambda)}|^2\,,\quad m^{-+}=(m^{+-})^*\;.
\end{eqnarray}

We choose $\lambda=+1$ for the initial fermion
\begin{eqnarray}
T_e=\sum_\lambda|m^+_{{\lambda}}|^2\,.
\end{eqnarray}
The cross sections $d\sigma_{n_1,n_2}$ of the process of type $2\to n_1+n_2$ with production of two jets
\begin{eqnarray}
a(p_1,\eta_1)+b(p_2,\eta_2) \to a_1(r_1\lambda_1)+\dots +a_{n_1}(r_{n_1},\lambda_{n_1})+
b_1(q_1,\sigma_1)+\dots +b_{n_2}(q_{n_2},\sigma_{n_2})\,,
\end{eqnarray}
where energy fractions $x_1,\dots x_{n_1},\sum x_i=1$ and transversal components of momenta
$\vec{r}_1,\dots \vec{r}_{n_1}$, $\sum\vec{r}_i=\vec{q}$ of jet $a$ and similar quantities
$y_i,\vec{q}_i,\sum y_i=1,\sum \vec{q}_i=-\vec{q}$ for the other jet $b$, have the form
\begin{eqnarray}
d\sigma_{22}=\frac{\alpha^4}{2^2\pi^4}T^{(1)}_2T^{(2)}_2
\frac{d^2q}{(\vec{q}^2)^2}d^2r_1d^2q_1\frac{dx_1dy_1}{x_1x_2y_1y_2}\, ,
\end{eqnarray}
\begin{eqnarray}
d\sigma_{23}=\frac{\alpha^5}{2^4\pi^6}T^{(1)}_2T^{(2)}_3
\frac{d^2q}{(\vec{q}^2)^2}d^2r_1d^2q_1d^2q_2\frac{dx_1dy_1dy_2}{x_1x_2y_1y_2y_3}\, ,
\end{eqnarray}
\begin{eqnarray}
d\sigma_{33}=\frac{\alpha^6}{2^6\pi^8}T^{(1)}_3T^{(2)}_3
\frac{d^2q}{(\vec{q}^2)^2}d^2q_1d^2q_2d^2r_1d^2r_2 %\\ \nonumber
\frac{dx_1dx_2dy_1dy_2}{x_1x_2x_3y_1y_2y_3}\;.
\end{eqnarray}

\subsection{Subprocesses $\gamma^*\gamma\to e^+e^-,\,\pi^+\pi^-$}

Let us consider first the contribution to the photon impact factor from the
lepton pair production subprocess
\begin{eqnarray}
\gamma(k_1,\eta)+\gamma^*(q)\to e^-(q_-,\lambda)+e^+(q_+,-\lambda).\nn
\end{eqnarray}
The matrix element of the subprocess has the form (we suppress the factor $4\pi \alpha$)
\begin{eqnarray}
m^{\eta \mu}_{1 \lambda}=
-\bar{u}_{\lambda}(q_-)\Big[\hat {\varepsilon}^\eta \frac{\hat{q}_- - \hat{k}_1}
{\kappa_{1-}}\gamma^{\mu}+ \gamma^\mu
\frac{-\hat{q}_+ + \hat{k}_1}{\kappa_{1+}}\hat{\varepsilon}^\eta\Big] v_{\lambda}(q_+)\,
,\;\bar{u}_\lambda=\bar{u}\, \omega_{-\lambda}, \, v_\lambda =\omega_{-\lambda} v\;.
\end{eqnarray}

We imply all the particles to be massless. A definite chiral state initial photon polarization vector
has the form \cite{B}
\begin{eqnarray}
\hat{\varepsilon}_1^\lambda = N_1[\hat{q}_-\hat{q}_+\hat{k}_1\omega_{-\lambda}-\hat{k}_1\hat{q}_-\hat{q}_+\omega_\lambda]\,,
\end{eqnarray}
where
\begin{eqnarray}
N_1^2=\frac{2}{s_1\kappa_+ \kappa_-}\,,\quad s_1=2q_+q_-\,, \quad \kappa_{1\pm}=2k_1q_\pm\;.
\end{eqnarray}
Chiral amplitudes $m^\eta_\lambda= (1/s)m_{1\lambda}^{\eta\mu} p_{2\mu}$ have the form
\begin{eqnarray}
m_{1+}^+=-\frac{N_1}{s}\bar{u}\hat{q_+}\hat{q} \hat{p}_2\omega_+v\;, \quad
m_{1-}^+=-\frac{N_1}{s}\bar{u}\hat{p}_2 \hat{q} \hat{q}_-\omega_-v\; , \\ \nonumber
m_{1-}^-=-\frac{N_1}{s}\bar{u}\hat{q}_+ \hat{q} \hat{p}_2\omega_-v\;, \quad
m_{1+}^-=-\frac{N_1}{s}\bar{u}\hat{p}_2 \hat{q} \hat{q}_-\omega_+v\;.
\end{eqnarray}
The elements of the spin-matrix $\cal{ M}$ in the case of lepton pair production are,
\begin{eqnarray}
m_{e^+e^-}^{++}=m_{e^+e^-}^{--}=\frac{2\vec{q}^2}{\vec{q}_+^2\vec{q}_-^2}x_+x_-(x_+^2+x_-^2)\;, \\ \nonumber
m_{e^+e^-}^{+-}=(m_{e^+e^-}^{-+})^*=-\frac{4\vec{q}^2}{\vec{q}_+^2\vec{q}_-^2}(x_+x_-)^2 e^{2i\theta}\;,
\end{eqnarray}
$x_\pm$ are the energy fractions carried out by pair components, $x_++x_-=1$ and $\theta$ is the angle
between two Euclidean vectors $\vec{q}=\vec{q}_-+\vec{q}_+$ and
$\vec{Q}=x_+\vec{q}_--x_-\vec{q}_+$.

In the case of charged pion pair production
\begin{eqnarray}
\gamma(p_1,e_1^\eta)+\gamma^*(q)\to \pi^+(q_+)+\pi^-(q_-)
\end{eqnarray}
we have
\begin{eqnarray}
m^\eta =\frac{1}{s}\varepsilon_{1\nu}^\eta p_2^\mu m_{\mu}^{\nu}=\frac{x_+}{p_1q_-}\varepsilon_1^\eta q_-+
\frac{x_-}{p_1q_+}\varepsilon_1^\eta q_+-\frac{2}{s}(\varepsilon_1^\eta p_2)\;.
\end{eqnarray}
Using the photon polarization vector written as
\begin{eqnarray}
\varepsilon_{1\mu}^\eta=N_1[(q_+p_1) q_{-\mu}-(q_-p_1) q_{+\mu}+i\eta\varepsilon_{\mu\alpha\beta\gamma}q_-^\alpha q_+^\beta p_1^\gamma]\,,
\end{eqnarray}
we obtain the chiral amplitude of the pion pair production process (we define $(p_1p_2q_-q_+)=
\epsilon_{\alpha\beta\gamma\delta}p_1^\alpha p_2^\beta q_-^\gamma q_+^\delta =(s/2)[\vec{q}_- \vec{q}_+]_z$)\;
\begin{eqnarray}
m^\eta=-N_1(\vec{Q}\vec{q}+i\eta[\vec{Q},\vec{q}]_z)=-N_1|\vec{q}|\,|\vec{Q}|e^{i\eta \theta}, \quad
\theta=\widehat{\vec{q}\vec{Q}}.
\end{eqnarray}
where we imply the $z$ axis direction along the photon 3-vector and use the
relation $[\vec{q}_-,\vec{q}_+]_z=[\vec{Q},\vec{q}]_z$.
For the pion chiral matrix we have
\begin{eqnarray}
m_{\pi^+\pi^-}^{++}=m_{\pi^+\pi^-}^{--}=\frac{2\vec{q}^2}{\vec{q}_+^2\vec{q}_-^2}(x_+x_-)^2\;, \\ \nonumber
m_{\pi^+\pi^-}^{+-}=(m_{\pi^+\pi^-}^{-+})^*=\frac{2\vec{q}^2}{\vec{q}_+^2\vec{q}_-^2}(x_+x_-)^2 e^{2i\theta}\;.
\end{eqnarray}
For the two-pair production process
\begin{eqnarray}
\gamma_1(p_1,\vec{\xi}_1)+\gamma_2(p_2,\vec{\xi}_2)\to a(q_-)+\bar{a}(q_+)+b(p_-)+\bar{b}(p_+)\;, \\ \nonumber
q_\pm=\alpha_\pm p_2+x_\pm p_1+q_{\pm\bot};\quad p_\pm=y_\pm p_2+\beta_\pm p_1+p_{\pm\bot}\;,
\end{eqnarray}
the differential cross section (assuming that the pair
$a\bar{a}$ moves along the photon 1 direction and the pair $b\bar{b}$ moves along the photon 2 direction)
has the form (22)
%
%d\sigma^{\gamma\gamma\to a\bar{a} b\bar{b}}=\frac{\alpha^4 T^{(1)}_2 T^{(2)}_2}{4\pi^4}
%\frac{d^2q d^2q_- d^2q_+}{\vec{q}_-^2\vec{p}_-^2(\vec{q}-\vec{q}_-)^2(\vec{q}+\vec{p}_-)^2}\;,
%\ea
with
\begin{eqnarray}
T^{(1)}=\frac{\vec{q}^2}{\vec{q}^2_+\vec{q}^2_-}(x_+x_-)^2
[1-\xi_3\cos(2\theta)+\xi_1\sin(2\theta)]\;,\quad \mathrm{for}\quad\pi^+,\pi^-\;, \\
T^{(1)}=\frac{\vec{q}^2}{\vec{q}^2_+\vec{q}^2_-}(x_+x_-)\{
x_+^2+x_-^2+2x_+x_-[\xi_3\cos(2\theta)+\xi_1\sin(2\theta)]\}\,\quad \mathrm{for}\quad e^+,e^-
\label{T1}
\end{eqnarray}
and the similar expression for $T^{(2)}$ \footnote{In paper \cite{BDGK} formula \ref{T1} contains a misprint
in the sign of $\xi_3^{(1,2)}$.}.
We remind that the formulae obtained are valid at a large, compared to masses of particles,
transverse component of jet particles
\begin{eqnarray}
\vec{q}_-^2\sim\vec{q}_+^2\sim\vec{p}_+^2\sim\vec{p}_-^2>>m^2,\quad \vec{q}_+=\vec{q}-\vec{q}_-;\quad
\vec{p}_+=-\vec{q}-\vec{p}_-,
\end{eqnarray}
and finite energy fractions $x_\pm\sim y_\pm\sim 1$, which corresponds to the
emission angles of jet particles $\theta_i=|\vec{q}_i|/(x_i\varepsilon)>>m/\varepsilon$
that are considerably larger than the mass to energy ratio.

\subsection{Subprocesses $\gamma^*\gamma\to e^+e^-\gamma,\pi^+\pi^-\gamma$}

Here and below for subprocesses of type $2 \to 3$ we restrict ourselves to calculating
the chiral amplitudes and checking their gauge invariance properties.

The subprocess
\begin{eqnarray}
\gamma(k,\lambda)+\gamma^*(q)\to e^+(q_+,-\lambda_-)+e^-(q_-,\lambda_-)+\gamma(k_1,\lambda_1)\;,\nn
\end{eqnarray}
is described by 6 FD. A standard calculation of chiral amplitudes $m^\lambda_{\lambda_1\lambda_-}$ leads to
\begin{eqnarray}
m^+_{++}=-\frac{s_1NN_1}{s}\bar{u}(q_-)\hat{q}_+\hat{q}\hat{p}_2\omega_+v(q_+)=(m^-_{--})^*\,,  \nonumber\\
m^+_{+-}=-\frac{s_1NN_1}{s}\bar{u}(q_-)\hat{p}_2\hat{q}\hat{q}_-\omega_-v(q_+)=(m^-_{-+})^*\,, \\ %\nonumber
m^+_{-+}=\frac{NN_1}{s}\bar{u}(q_-)A^+_{-+}\omega_+v(q_+)=(m^-_{+-})^*\,, \nonumber \\
m^+_{--}=\frac{NN_1}{s}\bar{u}(q_-)A^+_{--}\omega_-v(q_+)=(m^-_{++})^*\,, \nonumber
\end{eqnarray}
with $A^+_{--}(k,k_1)=A^+_{-+}(-k_1,-k)$
\begin{eqnarray}
N^2=\frac{2}{s_1\kappa_-\kappa_+},\quad N_1^2=\frac{2}{s_1\kappa_{1+}\kappa_{1-}},\quad s_1=2q_+q_-,\quad
\kappa_\pm=2kq_\pm,\quad \kappa_{1\pm}=2k_1q_\pm
\end{eqnarray}
and a rather cumbersome expression for $A^+_{-+}$
\begin{eqnarray} \label{exA}
A^+_{-+}=\frac{s_1}{(q_+-q)^2}\hat{k}\hat{q}_+\hat{k}_1(-\hat{q}_++\hat{q})\hat{p}_2-\hat{q}_+(\hat{q}_--\hat{k})\hat{p}_2(\hat{q}_++\hat{k}_1)\hat{q}_-\nn \\
-\frac{s_1}{(q_--q)^2}\hat{p}_2(\hat{q}_--\hat{q})\hat{k}\hat{q}_-\hat{k}_1\;.
\end{eqnarray}
Substituting
$$ \hat{p}_2\approx\frac{1}{\alpha}(\hat{q}-\hat{q}_\bot)=\frac{s}{s\alpha}
[\hat{q}_++\hat{k_1}+(\hat{q}_--\hat{k})-\hat{q}_\bot]\;, $$
in the second term of r.h.s. (\ref{exA})  we have
\begin{eqnarray}
A^+_{-+}=-ss_1\kappa_{1+}[\frac{x_+}{(q_+-q)^2}+\frac{1}{s\alpha}]\hat{k}-
ss_1\kappa_-[\frac{x_-}{(q_--q)^2}+\frac{1}{s\alpha}]\hat{k}_1 \nn\\
+\frac{s_1}{(q_+-q)^2}\hat{k}\hat{q}_+\hat{k}_1\hat{q}_\bot\hat{p}_2+
\frac{s_1}{(q_--q)^2}\hat{p}_2\hat{q}_\bot\hat{k}\hat{q}_-\hat{k}_1+
\frac{s}{s\alpha}\hat{q}_+(\hat{q}_--\hat{k})\hat{q}_\bot(\hat{q}_++\hat{k}_1)\hat{q}_-\;,
\end{eqnarray}
with
\begin{eqnarray}
(q_\pm-q)^2=-\vec{q}^2+2\vec{q}\vec{q}_\pm-s\alpha x_\pm,\quad
s\alpha=\frac{\vec{k}_1^2}{x_1}+\frac{\vec{q}_-^2}{x_-}+\frac{\vec{q}_+^2}{x_+}\,,\\ \nonumber
x_1+x_-+x_+=1, \quad
\kappa_\pm=\frac{\vec{q}_\pm^2}{x_\pm},\quad
\kappa_{1\pm}=\frac{1}{x_1x_\pm}(x_1\vec{q}_\pm-x_\pm\vec{k}_1)^2\,.
\end{eqnarray}
A gauge property ( the chiral amplitudes must vanish as $\vec{q}\to 0$) can be seen explicitly.

A further procedure of constructing the chiral matrix is straightforward and can be performed in terms of
simple traces. We will not touch it here.

Consider the subprocess
\begin{eqnarray}
\gamma(k,\lambda)+\gamma^*(q)\to\pi^+(q_+)+\pi^-(q_-)+\gamma(k_1,\lambda_1)\;.\nn
\end{eqnarray}
There are 12 FD describing rather a cumbersome expression for the matrix element.
It can be considerably simplified when using the modified expressions for the photon
polarization vectors in the form \cite{S}
\begin{eqnarray}
\varepsilon_\mu^\lambda(k)=\frac{N}{2}Sp\gamma_\mu \hat{q}_-\hat{q}_+\hat{k}\omega_\lambda\,, \qquad
\varepsilon_\mu^{\lambda_1}(k_1)=\frac{N_1}{2}Sp\gamma_\mu \hat{q}_-\hat{q}_+\hat{k}_1\omega_\lambda
\end{eqnarray}
with the same expressions for $N,N_1$ as in the case of the $\gamma\gamma^*\to e^+e^-\gamma$ subprocess.
Polarization vectors chosen in such a form satisfy Lorenz condition $\varepsilon(k)k=0,
\varepsilon(k_1)k_1=0$ and gauge condition $\varepsilon(k)q_-=\varepsilon(k_1)q_-=0$.

The matrix element has the (we  lost the Bose symmetry at this stage) form
\begin{eqnarray}
m^\lambda_{\lambda_1}&=&\frac{1}{s}p_2^\rho \varepsilon^{\mu}(k)\varepsilon_1^{*\sigma}(k_1)O_{\rho\mu\sigma}\nn\\
&=&\frac{4x_-}{(q_--q)^2}\Big[\frac{(\varepsilon_1q_+)(\varepsilon q)}{\kappa_{1+}}-
\frac{(\varepsilon_1q)(\varepsilon q_+)}{\kappa_+}\Big]+\frac{4(\varepsilon p_2)(\varepsilon_1q_+)}{s\kappa_{1+}}-
\frac{4(\varepsilon_1p_2)(\varepsilon q_+)}{s\kappa_+}\nn \\
&+&2(\varepsilon\varepsilon_1)\Big[\frac{x_+}{(q_+-q)^2}-\frac{x_-}{(q_--q)^2}\Big]\,,
\end{eqnarray}
where we imply $\varepsilon=\varepsilon^\lambda$, $\varepsilon_1=\varepsilon_1^{\lambda_1}$
and $x_\pm=2p_2q_\pm/s$, $x_1=2p_2k_1/s$ where $x_++x_-+x_1=1$.

For $\lambda_1=\lambda$ we have
\begin{eqnarray}
m^\lambda_\lambda=s_1NN_1[A_1+i\lambda B_1]\,,\quad A_1=-\vec{Q}\vec{q},\quad B_1=[\vec{Q}\vec{q}]_z\,.
\end{eqnarray}
For the case of opposite chiralities we have
\begin{eqnarray}
m^\lambda_{-\lambda}&=&s_1NN_1[A+i\lambda B], \\ \nonumber
A&=&-\vec{Q}\vec{q}+\frac{1}{2x_1x_-x_+}[\vec{Q}^2\vec{k}_1^2-\vec{q}_-^2(x_1\vec{q}_+-x_+\vec{k}_1)^2-
\vec{q}_+^2(x_1\vec{q}_--x_-\vec{k}_1)^2]\times \\ \nonumber
&&\Big(\frac{x_+}{(q_+-q)^2}-\frac{x_-}{(q_--q)^2}\Big), \\ \nonumber
B&=&\Big(\frac{x_+}{(q_+-q)^2}+\frac{x_-}{(q_--q)^2}\Big)
(s\alpha[\vec{q}_-\vec{q}_+]_z-s\alpha_-[\vec{q}\vec{q}_+]_z+
s\alpha_+[\vec{q}\vec{q}_-]_z)+
2[\vec{q}_-\vec{q}_+]_z-[\vec{Q}\vec{q}]_z\,, \\ \nonumber
&&s\alpha_\pm=\frac{\vec{q}_\pm^2}{x_\pm}\,, \quad
s\alpha=\frac{\vec{k}_1^2}{x_1}+s\alpha_++s\alpha_-\,.
\end{eqnarray}
We can see that the Bose-symmetry is restored.
%The matrix elements of the chiral matrix are
%\ba
%m^{++}=(s_1NN_1)^2[A^2+B^2+A_1^2+B_1^2]=m^{--}; \\ \nonumber
%m^{+-}=
%(m^{-+})^*=2(s_1NN_1)^2[AA_1+iBB_1(A_1B-B_1A)]
%\ea
%and the conversion with Stoke's matrix is
%\ba
%Tr(\rho{\cal M})=\frac{4}{\kappa_-\kappa_+\kappa_{ 1 - }\kappa_{ 1 + }} \\ \nonumber
%[A^2+A_1^2+B^2+B_1^2-2\xi_3(AA_1+BB_1)+
%\xi_1(A_1B-B_1A)].
%\ea

\subsection{ Subprocesses $e\gamma^*\to e\gamma;e+\gamma+\gamma$}

Consider first the Compton subprocess \footnote{The case of real initial photons
was considered in paper \cite{egeg}. }
\begin{eqnarray}
\gamma^*(q)+e(p,\lambda_1)\to\gamma(k,\lambda)+e(p',\lambda_1)\,.\nn
\end{eqnarray}
For the chiral matrix elements we have (we chose $\lambda_1=+1)$
\begin{eqnarray}
m^+_\lambda=\frac{N}{s}\bar{u}(p')[-\hat{p}\omega_\lambda(\hat{p}'+\hat{k})\hat{p}_2
-\hat{p}_2(\hat{p}-\hat{k})\hat{p}'\omega_{-\lambda}]\omega_+u(p)\,, \\ \nonumber
m^+_+=-\frac{N}{s}\bar{u}(p')\hat{p}\hat{q}\hat{p}_2\omega_+u(p)\,,\quad
m^+_-=-\frac{N}{s}\bar{u}(p')\hat{p}_2\hat{q}\hat{p}'\omega_+u(p)\,.
\end{eqnarray}
The sum of modulo square of the matrix elements is
\begin{eqnarray}
T_e=\sum_\lambda|m^+_\lambda|^2=2\frac{\vec{q}^2}{\kappa\kappa'}[1+(1-x)^2]\,,
\end{eqnarray}
with
\begin{eqnarray}
\kappa=2kp=\frac{\vec{k}^2}{x},\quad \kappa'=2kp'=\frac{1}{x(1-x)}(\vec{p}'x-\vec{k}(1-x))^2,
\end{eqnarray}
and $x=2kp_2/2p_1p_2,1-x$ are the energy fractions of photon and electron in the final state.

Consider now the double Compton subprocess (see Fig. 3a)
\begin{eqnarray}
e(p,\eta)+\gamma^*(q)\to e(p',\eta)+\gamma(k_1,\lambda_1)+\gamma(k_2,\lambda_2).
\end{eqnarray}
The chiral matrix elements $m^{\eta}_{\lambda_1\lambda_2}$ are
\begin{eqnarray}
m^+_{++}=(m^-_{--})^*=-\frac{s_1N_1N_2}{s}\bar{u}(p')\hat{p}\hat{q}\hat{p}_2\omega_+u(p); \\ \nonumber
m^+_{--}=(m^-_{++})^*=-\frac{s_1N_1N_2}{s}\bar{u}(p')\hat{p}_2\hat{q}\hat{p}'\omega_+u(p); \\ \nonumber
m^+_{+-}=(m^-_{-+})^*=\frac{N_1N_2}{s}\bar{u}(p')A^+_{+-}\omega_+u(p); \\ \nonumber
m^+_{-+}=(m^-_{+-})^*=\frac{N_1N_2}{s}\bar{u}(p')A^+_{-+}\omega_+u(p),
\end{eqnarray}
with $A^+_{-+}(k_1,k_2)=A^+_{+-}(k_2,k_1)$ and
\begin{eqnarray}
A^+_{+-}(k_1,k_2)=\frac{s_1}{(p'-q)^2}\hat{p}_2(\hat{p}'-\hat{q})\hat{k}_1\hat{p}'\hat{k}_2+\hat{p}(\hat{p}'+\hat{k}_1)\hat{p}_2(\hat{p}-\hat{k}_2)\hat{p}'+\\ \nonumber
\frac{s_1}{(p+q)^2}\hat{k}_1\hat{p}\hat{k}_2(\hat{p}+\hat{q})\hat{p}_2,
\end{eqnarray}
with
\begin{eqnarray}
s_1=2pp',\quad N_i^2=\frac{2}{s_1\kappa_i\kappa_i'},\quad \kappa_i=2pk_i,\quad \kappa_i'=2p'k_i.
\end{eqnarray}
To see the gauge invariance property of two last amplitudes, we make a substitution
$p_2=(q-q_\bot)/\alpha_q$ in the second term of r.h.s. and arrive at the form
\begin{eqnarray}
A^+_{+-}(k_1,k_2)=ss_1\kappa_1'(\frac{x'}{(p'-q)^2}+\frac{1}{s\alpha_q})\hat{k}_2 +
ss_1\kappa_2(\frac{1}{(p+q)^2}-\frac{1}{s\alpha_q})\hat{k}_1+\nn \\
\frac{s_1}{(p+q)^2}\hat{k}_1\hat{p}\hat{k}_2\hat{q}_\bot\hat{p}_2-
\frac{s_1}{(p'-q)^2}\hat{p}_2\hat{q}_\bot\hat{k}_1\hat{p}'\hat{k}_2 -\hat{p}(\hat{p}'+\hat{k}_1)\hat{q}_\bot(\hat{p}-\hat{k}_2)\hat{p}'\frac{s}{s\alpha_q}.
\end{eqnarray}
We can verify that this expression turns to zero at $\vec{q}=0$. Really, we can use
\begin{eqnarray}
(p'-q)^2 = -\vec{q}^{~2}+2\vec{p}^{~'}\vec{q}-sx'\alpha_q,\quad (p+q)^2=-\vec{q}^{~2}+s\alpha_q, \\ \nonumber
\alpha_q=\alpha'+\alpha_1+\alpha_2,\quad x'+x_1+x_2=1,\quad
s\alpha'=\frac{(s\vec{p}^{~'})^2}{x'},\quad s\alpha_i=\frac{\vec{k}_i^2}{x_i},\\ \nonumber \kappa_i=s\alpha_i, \quad
\kappa'_i=\frac{1}{x'x_i}(\vec{k}_ix'-\vec{p}^{~'}x_i)^2.
\end{eqnarray}
A further strategy is similar to the one mentioned above (45).

\subsection{Subprocesses $e\gamma^*\to e \pi^+\pi^-,e\mu^+\mu^-$ }

The matrix element of the pion pair production subprocess
\begin{eqnarray}
e(p,\eta)+\gamma^*(q)\to\pi^+(q_+)+\pi^-(q_-)+e(p',\eta)\nn
\end{eqnarray}
can be written in the form
\begin{eqnarray}
m^\eta=\bar{u}(p')[\hat{B}+\hat{D}]\omega_\eta u(p),
\end{eqnarray}
where bremsstrahlung mechanism contribution is (see Fig.3b)
\begin{eqnarray}
\hat{B}=\frac{1}{q_1^2}\Big[B\hat{q}_1+\frac{1}{s(p+q)^2}\hat{q}_1\hat{q}\hat{p}_2-
\frac{1}{s(p'-q)^2}\hat{p}_2\hat{q}\hat{q}_1\Big],\quad q_1=q_++q_-,\quad q_2=p^{'}-p_1; \\ \nonumber
\hat{D}=\frac{1}{q_2^2}\Big[D(2\hat{q}_- +\hat{q}_2)-2\frac{x_-}{(q-q_-)^2}\hat{q}_\bot
+\frac{2(\vec{q}^{~2}-2\vec{q}\,\vec{q}_-)}{s(q-q_-)^2}\hat{p}_2\Big].
\end{eqnarray}
For the squares of modulo of the chiral amplitudes, which enter in (23,24), we have
\begin{eqnarray}
T^{(\pi)}_3=|m^+|^2=Sp(\hat{p}'(\hat{B}+\hat{D})\hat{p}(\tilde{B}+\tilde{D})\omega_+)\,,
\end{eqnarray}
with $B$ and $D$ specified below (\ref{BD}).

For the subprocess of the muon pair production
\begin{eqnarray}
e(p,\eta)+\gamma^*(q)\to\mu^+(q_+)+\mu^-(q_-)+e(p',\eta).
\end{eqnarray}
The bremsstrahlung and two-photon mechanisms must be taken into account (see Fig.3b,c)
\begin{eqnarray}
m^+_\lambda=\frac{1}{q_1^2}\bar{u}(p')B_\mu\omega_+ u(p)\times\bar{u}(q_-)\gamma^\mu\omega_\lambda v(q_+)
+\frac{1}{q_2^2}\bar{u}(p')\gamma_\nu \omega_+ u(p)\bar{u}(q_-)D_\nu\omega_\lambda v(q_+)\,,
\end{eqnarray}
with double photon mechanism contribution (not considered in paper \cite{C})
\begin{eqnarray}
D_\nu =D\gamma_\nu+\frac{1}{s(q-q_+)^2}\gamma_\nu\hat{q}\hat{p}_2-\frac{1}{s(q-q_-)^2}\hat{p}_2\hat{q}\gamma_\nu\nn
\end{eqnarray}
and bremsstrahlung mechanism one
\begin{eqnarray}
B_\mu=B\gamma_\mu-\frac{1}{s(p'-q)^2}\hat{p}_2\hat{q}\gamma_\mu+\frac{1}{s(p+q)^2}\gamma_\mu\hat{q}\hat{p}_2\nn\;,
\end{eqnarray}
with
\begin{eqnarray}   \label{BD}
B=\frac{x^{'}}{(p'-q)^2}+\frac{1}{(p+q)^2}, \quad D=\frac{x_-}{(q_--q)^2}-\frac{x_+}{(q-q_+)^2}\;,\nn\\
 x_\pm=\frac{2p_2q_\pm}{s}, \quad x^{'}=\frac{2p_2p^{'}}{s},\quad x_++x_-+x^{'}=1\,. \nn
\end{eqnarray}
To perform the conversion in the Lorentz indices $\mu,\nu$ in (66), one can use the
projection operators. For the case of equal chiralities  $\eta=\lambda=+1$ we choose the projection
operator as
\begin{eqnarray}
P_+=\frac{\bar{u}(p)\hat{q}_+\omega_+u(q_-)}{\bar{u}(p)\hat{q}_+\omega_+u(q_-)}\;.
\end{eqnarray}
Inserting it and using the relation $\omega_+u(p)\bar{u}(p)=\omega_+\hat{p}$, we obtain
\begin{eqnarray}
m_+^+= \frac{-2}{\bar{u}(p)\hat{q}_+\omega_+ u(q_-)} \bar{u}(p')\Big[\left(\frac{D}{q_2^2}
+\frac{B}{q_1^2}\right)\hat{q}_-\hat{q}_+\hat{p}+
\frac{\hat{q}_- \hat{q}_+ \hat{p} \hat{q}_\bot \hat{p}_2}{s}\left(\frac{1}{q_2^2 (q-q_+)^2}-\frac{1}{q_1^2 (p+q)^2}\right) \nn\\
+\frac{\hat{p}_2\hat{q}_\bot \hat{q}_-\hat{q}_+\hat{p}}{s}\left(\frac{1}{q_2^2 (q_-- q)^2}
-\frac{1}{q_1^2 (p'-q)^2}\right)\Big]\omega_+ v(q_+)\nn\\=
\frac{-2}{\bar{u}(p)\hat{q}_+\omega_+ u(q)_-} \bar{u}(p')A^+_+\omega_+v(q_+)\;.\qquad
\end{eqnarray}
For the case of opposite chiralities $\eta=-\lambda=+1$ we use the projection operator
\begin{eqnarray}
P_-=\frac{\bar{u}(p)\omega_-u(q_-)}{\bar{u}(p)\omega_-u(q_-)}\;.
\end{eqnarray}

Similar calculations lead to the result
\begin{eqnarray}
m_-^+=\frac{2}{\bar{u}(p)\omega_- u(q_-)}\bar{u}(p')\Big[\left(\frac{D}{q_2^2}+\frac{B}{q_1^2}\right)2(pq_-)+
2\frac{\hat{p}\hat{q}_-\hat{q}_\bot \hat{p}_2}{s}\left(\frac{1}{q_2^2(q-q_+)^2}+\frac{1}{q_1^2(p_1 -q_-)^2}\right)\nn\\
-\frac{\hat{p}\hat{q}_\bot \hat{p}_2\hat{q}_-}{s}\left(\frac{1}{q_2^2(q-q_-)^2}+\frac{1}{q_1^2(p+q)^2}\right)-
\frac{\hat{q}_- \hat{p}_2 \hat{q}_\bot \hat{p}}{s}\left(\frac{1}{q_2^2(q-q_-)^2}+\frac{1}{q_1^2(p+q)^2}\right)\Big]\omega_- v(q_+)\nn \\
=\frac{2}{\bar{u}(p)\omega_- u(q_-)}\bar{u}(p')A^+_-\omega_-v(q_+)\;.\qquad
\end{eqnarray}
The property of $A^+_+, A^+_-$ tending to zero as $|\vec{q}|\to 0$ is explicitly seen  from (71, 72).

For the sum of squares of chiral amplitudes entering (23,24), one has
\begin{eqnarray}
T^{(\mu)}_3=\sum|m^+_\lambda|^2= \frac{1}{(pq_+)(q_-q_+)}Sp(\hat{p}'A^+_+\hat{q}_+\tilde{A}^+_+\omega_+)
+\frac{2}{pq_-}Sp(\hat{p}'A^+_-\hat{q}_+\tilde{A}^+_-\omega_+)\;.
\end{eqnarray}
A further strategy is straightforward.

\subsection{Subprocess $e\gamma^*\to ee\bar{e}$}

Kinematics of subprocess is defined as
\begin{eqnarray}
e(p,l_p)+\gamma^*(q)\to e(p_1,l_1)+e(p_2,l_2)+\bar{e}(p_+,t),\nn
\end{eqnarray}
with $l_i,t=\pm$ the chiralities of initial and final fermions. Without loss of
generality we can put below $l_p=+$. For the sum on chiral states of the modulo
square of relevant matrix element we obtain:
\begin{eqnarray}
\sum |M^{l_p}_{l_1l_2t}|^2=2[|M^+_{++-}|^2+|M^+_{+-+}|^2+|M^+_{-++}|^2].
\end{eqnarray}
Eight Feynman diagrams are relevant which form 4 gauge-invariant sets of amplitudes:
\begin{eqnarray}
M^+_{l_1l_2t}=(4\pi\alpha)^{\frac{3}{2}}\left(-\frac{1}{s_1}\right)(\delta_{l_1,+}
\delta_{t,-l_2}[\bar{u}^{l_2}(p_2)\gamma_{\lambda}v^t(p_+)\bar{u}^{l_1}A_\lambda
u^+(p)\nn\\+\bar{u}^{l_1}(p_1)\gamma_{\sigma}\bar{u}^+(p)\bar{u}^{l_2}B_\sigma v^t(p_+)]\nn\\
+\delta_{l_2,+}\delta_{t,-l_1}[ \bar{u}^{l_1}(p_1)\gamma_{\eta}v^t(p_+)
\bar{u}^{l_2}(p_2)D_\eta u^+(p)+\bar{u}^{l_2}(p_2)\gamma_{\delta}\bar{u}^+(p)
\bar{u}^{l_1}(p1)C_\delta v^t(p_+)]).
\end{eqnarray}
Applying projection operators to provide the conversion on vector indices we have
\begin{eqnarray}
| M^+_{++-}|^2=\frac{(4\pi\alpha)^3}{2s^2_1 pp_+}
\left[\frac{1}{p_2p_+}\frac{1}{4}{\rm Sp}\hat{p}_1m^{(1)}_{++-}\hat{p}_+(m^{(1)}_{++-})^+\right.\nn\qquad\\
\left.+\frac{1}{p_+p_1}\frac{1}{4}{\rm Sp} \hat{p}_2m^{(2)}_{++-}\hat{p}_+(m^{(2)}_{++-})^+\frac{1}{p_1p_+ p_2p_+}
\frac{1}{4}{\rm Sp} \hat{p}_1m^{(1)}_{++-}\hat{p}_+(m^{(2)}_{++-})^+\hat{p}_2\hat{p}_+\right];\qquad\\
| M^+_{+-+}|^2=\frac{(4\pi\alpha)^3}{2s^2_1 pp_2}\frac{1}{4}{\rm Sp}\hat{p}_1m_{+-+}\hat{p}_+(m_{+-+})^+,\quad
| M^+_{-++}|^2=\frac{(4\pi\alpha)^3}{2s^2_1 pp_1}\frac{1}{4}{\rm Sp}\hat{p}_2m_{-++}\hat{p}_+(m_{-++})^+;\nn
\end{eqnarray}
with
\begin{eqnarray}
m_{+-+}=\gamma_\sigma \hat{p}\hat{p}_2B_\sigma+A_\lambda\hat{p}\hat{p}_2\gamma_\lambda,\qquad
m_{-++}=\gamma_\delta \hat{p}\hat{p}_1C_\delta+D_\eta\hat{p}\hat{p}_1\gamma_\eta,\\
m^{(1)}_{++-}=A_\lambda\hat{p}\hat{p}_+\hat{p}_2\gamma_\lambda+\gamma_\sigma\hat{p}\hat{p}_+\hat{p}_2B_\sigma,\qquad
m^{(2)}_{++-}=\gamma_\sigma\hat{p}\hat{p}_+\hat{p}_1C_\sigma+D_\eta \hat{p}\hat{p}_+\hat{p}_1\gamma_\eta,\nn\\
A_\lambda=\frac{\hat{q}_\bot(\hat{p}_1-\hat{q})\gamma_\lambda}{(p_1-q)^2}+\frac{\gamma_\lambda(\hat{p}+\hat{q})\hat{q}_\bot}{(p+q)^2}\qquad
B_\sigma=\frac{\hat{q}_\bot(\hat{p}_2-\hat{q})\gamma_\sigma}{(p_2-q)^2}+\frac{\gamma_\sigma(\hat{q}-\hat{p}_+)\hat{q}_\bot}{(p_+-q)^2}\\
C_\sigma=\frac{\hat{q}_\bot(\hat{p}_1-\hat{q})\gamma_\sigma}{(p_1-q)^2}+\frac{\gamma_\sigma(\hat{q}-\hat{p}_+)\hat{q}_\bot}{(q-p_+)^2}\qquad
D_\eta=\frac{\hat{q}_\bot(\hat{p}_2-\hat{q})\gamma_\eta}{(p_2-q)^2}+\frac{\gamma_\eta(\hat{p}+\hat{q})\hat{q}_\bot}{(p+q)^2}.\nn
\end{eqnarray}

\subsection*{Conclusion}

In our paper \cite{BDGK}, we wrote down the explicit expressions for the spin
matrix elements ${\cal{M}}_{ij}$ for subprocesses of the type $2\to
2$, which are reviewed here. For the subprocesses of the type $2\to 3$, we formulated the
algorithm of the calculation of spin matrix elements. We
considered all possibilities of pair creation in the mentioned
subprocesses as they were not completely considered in recent a
work \cite{C}. The gauge condition ${\cal{M}}_{ij}(q)\to
0$ for $|\vec{q}|\to 0$ is explicitly fulfilled in all cases. The
subprocesses with the pions in the final state were also
considered in the paper for the first time.

Radiative corrections to chiral amplitude was calculated only for some subprocesses
of type $2\to 2$ \cite{BBGK05}.

The magnitude of the cross sections (22-24) is of the order $\alpha^n/\mu^2\gg \alpha^n/s, n=4, 5, 6$ where
$\mu^2=\max(s_1, s_2)$ is large enough to be measured, and does not depend on $s$. The strategy of calculation
of cross section, using the helicity amplitudes of subprocesses $2\to 3$, is described above and can be implemented
to numerical programs which take into account details of experiments.

\section*{Acknowledgements}
This work work was supported by Russian Foundation for Basic Research, grant No. 03-02-17077 .
On of us(M.G.) acknowledges the support at Byelorussian Foundation for Basic Research, grant No. F03-183.


\begin{thebibliography}{99}
\bibitem{NLC} Zeroth-Order Design Report for the Next Linear Collider,
Report No. SLAC-474 (1996); JLC Design Study, KEK Report 97-1 (1997); R.D. Heuer et al., Technical
Design Report of 500 GeV Electron Positron Collider with Integrated X-Ray Facility (DESY 2001-011, March 2001); Physics Potential and Development of
$\mu^+ \mu^-$ Colliders, Ed. D. Cline (AIP Conference Proceedings 441 (1997) )

\bibitem{C}
C. Carimalo, A. Schiller and V.G. Serbo, Eur. Phys. J. C {\bf 23},633 (2002)

\bibitem{BFKK} V.N. Baier, V.S. Fadin, V.A. Khoze and E.A. Kuraev, Phys. Rep. C {\bf 78} N3 293(1982)

\bibitem{RKT}
V. Berestetsky, E. Lifshits and L. Pitaevski, Quantum Electrodynamics. (Nauka,  Moscow 1989)

\bibitem{B}
F. A. Berends et. al, Nucl. Phys. B{\bf 206}, 53 (1982)

\bibitem {BDGK} E. Barto\v{s} et. all,
% A.-Z. Dubnic\v{c}kov\'{a}, M.V. Galynskii, E.A. Kuraev,
Nucl. Phys. B{\bf 676}, 481 (2004)

\bibitem {egeg} E. Barto\v{s} et. all,
% M.V. Galynskii, S.R. Gevorkyan and E.A. Kuraev,
Nucl. Phys. B{\bf 676}, 390 (2004)


\bibitem{S}V.I. Borodulin, R.N. Rogalev, S.R. Slabospitsky (Serpukhov, IHEP), CORE: Compendium of relations: Version 2.1.
. Preprint IFVE-95-90, IHEP-95-90, Jul 1995. 108pp. e-Print Archive: hep-ph/9507456;
Private communication of S. Sikach (Minsk, Institute of Physics BAS, Belarus).

\bibitem{BBGK05}
V. Bytev, E. Bartos, M. Galynskii, and E. Kuraev,
ZhETP, V. {\bf 130} 259(2006)


\end{thebibliography}
\end{document}